\documentclass[pdflatex,sn-nature]{sn-jnl}

\usepackage{graphicx}%
\usepackage{multirow}%
\usepackage{amsmath,amssymb,amsfonts}%
\usepackage{amsthm}%
\usepackage{mathrsfs}%
\usepackage[title]{appendix}%
\usepackage{xcolor}%
\usepackage{textcomp}%
\usepackage{manyfoot}%
\usepackage{booktabs}%
\usepackage{algorithm}%
\usepackage{algorithmicx}%
\usepackage{algpseudocode}%
\usepackage{listings}%
\usepackage{siunitx}%

\begin{document}

\title[Article Title]{Entanglement-driven responses through multiscale 3D-printed knits}


\author[1]{\fnm{Bradley} \sur{Cline}}\email{bscline@cougarnet.uh.edu}

\author[1]{\fnm{Catherine} \sur{Bai}}\email{catbai@seas.upenn.edu}

\author[2]{\fnm{Sehui} \sur{Jeong}}\email{sehui@stanford.edu}

\author[3]{\fnm{Ling} \sur{Xu}}\email{xulingr@mit.edu}

\author[1]{\fnm{Yue} \sur{Wang}}\email{ywang298@cougarnet.uh.edu}

\author[3]{\fnm{James U.} \sur{Surjadi}}\email{jsurjadi@mit.edu}

\author[3]{\fnm{Carlos M.} \sur{Portela}}\email{cportela@mit.edu}

\author*[1]{\fnm{Tian} \sur{Chen}}\email{tianchen@uh.edu}

\affil*[1]{\orgdiv{Department of Mechanical and Aerospace Engineering}, \orgname{University of Houston}, \orgaddress{\street{4226 Martin Luther King Blvd}, \city{Houston}, \postcode{77002}, \state{TX}, \country{USA}}}

\affil[2]{\orgdiv{Department of Mechanical Engineering}, \orgname{Stanford University}, \orgaddress{\street{440 Escondido Mall}, \city{Stanford}, \postcode{94305}, \state{CA}, \country{USA}}}

\affil[3]{\orgdiv{Department of Mechanical Engineering}, \orgname{Massachusetts Institute of Technology}, \orgaddress{\street{77 Massachusetts Ave}, \city{Cambridge}, \postcode{02139}, \state{MA}, \country{USA}}}


\abstract{
For their resilience and toughness, filamentous entanglements are ubiquitous in both natural and engineered systems across length scales, from polymer-chain- to collagen-networks and from cable-net structures to forest canopies~\cite{burla2019mechanical,wu1989chain,jansen2018role,seymour1989storm,zhang2017dynamic}. 
Textiles are an everyday manifestation of filamentous entanglement: the remarkable resilience and toughness in knitted fabrics arise predominately from the topology of interlooped yarns~\cite{ding2024unravelling,tajiri2025curling}. 
Yet most architected materials do not exploit entanglement as a design primitive, and industrial knitting fixes a narrow set of patterns for manufacturability~\cite{singal2024programming}. Additive manufacturing has recently enabled interlocking structures such as chainmail, knot and woven assemblies, hinting at broader possibilities for entangled architectures~\cite{zhou20253d,moestopo2023knots,wirth20233d}.
The general challenge is to treat knitting itself as a three-dimensional architected material with predictable and tunable mechanics across scales. Here, we show that knitted architectures fabricated additively can be recast as periodic entangled solids whose responses are both fabric-like and programmable. 
We reproduce the characteristic behavior of conventional planar knits and extend knitting into the third dimension by interlooping along three orthogonal directions, yielding volumetric knits whose stiffness and dissipation are tuned by prescribed pre-strain. 
We propose a simple scaling that unifies the responses across stitch geometries and constituent materials~\cite{wang2021structured,mahadevan2024knitting,gonzalez2024pulling}. Further, we realize the same topology from centimeter to micrometer scales, culminating in the fabrication of what is, to our knowledge, the smallest knitted structure ever made. 
By demonstrating 3D-printed knits can be interpreted both as a traditional fabric comprised of a single yarn, as well as a novel architected material with defined periodicity, this work establishes the dual nature of entangled filaments and paves the way towards a new form of material architectures with high degrees of entanglement~\cite{sanchez20233d}.}




\maketitle

\section{Introduction}\label{sec1}

The industrial revolution transformed textile manufacturing, automating the production of a narrow set of knit topologies optimized for speed and uniformity~\cite{earnshaw1986lace,liu2019mechanism}. This efficiency came at the cost of geometric and topological freedom: the ways in which yarns can loop and intertwine cannot be freely altered~\cite{ding2023unravelling}. As a result, knit patterns are seldom designed with targeted physical behavior in mind. Yet, fabrics created by interlooping a single contiguous yarn are remarkably resilient and tough: they can undergo large reversible deformations, dissipate energy through frictional sliding, and tough against damage accumulated over repeated use. Similar to other highly entangled systems, the topological interactions between the filaments dictate the physical properties alongside the constituent material of the filaments themselves~\cite{gonzalez2024pulling,chen2011overview,roylance1979penetration,engel2007creation}. 

Recent works aim to correlate geometry, friction, and contact mechanics to the anisotropic and hysteretic responses of fabrics~\cite{ding2024unravelling,gonzalez2024pulling,mahadevan2024knitting}. However, such studies have remained confined to the limited design space afforded by traditional knitting machines. In parallel, the field of architected materials has demonstrated how the geometry of the microstructure can be designed to achieve physical behaviors unattainable in homogeneous solids, but these rarely exploit the dense entanglement and sliding interactions that define textiles. Bridging these domains offers a new opportunity: to merge the periodicity and tunability of architected materials with the entanglement and energy dissipation intrinsic to fabrics, yielding a new class of matter whose mechanics arise from both topology and composition.  

Here, we show that knitting, long confined to two-dimensional textiles, can be reimagined as a general strategy for designing three-dimensional entangled material architectures. By formulating a geometrically exact description of each stitch and using 3D printing, we create planar and volumetric knits with tunable loop parameters that directly control stiffness, strength, and energy dissipation. These printed knits faithfully reproduce the nonlinear, anisotropic, and hysteretic behavior of conventional fabrics. We uncover a universal scaling law that collapses the stress–strain responses of both traditional and printed knits onto a single master curve, revealing an entanglement-governed relationship independent of material. Extending knitting into three orthogonal directions, we realize volumetric knits with programmable coupling between orthogonal strains and demonstrate that this framework holds across scales by microfabricating knits with loop dimensions on the order of microns. Together, these results establish 3D printing knits as a new paradigm in architected materials, where programmable mechanical behavior emerges from controlled filament entanglement.

\begin{figure}[ht!]
    \centering
\includegraphics[width=\textwidth]{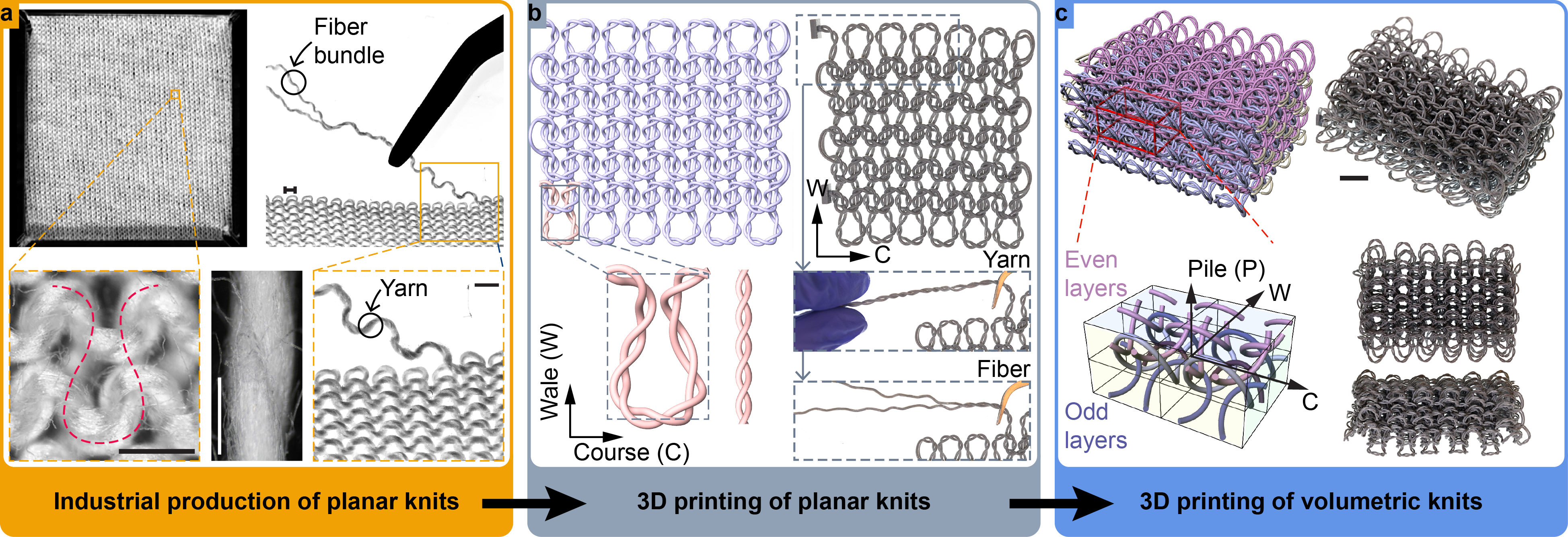}
    \caption{\textbf{3D printed knit material architectures.} \textbf{a}, A traditional Stockinette knit with a 2-ply cotton yarn. Unraveling of the knit shows individual yarn thread and fiber-plys. The scale bar is \SI{1}{\milli\metre}. \textbf{b}, Knit with the same 2-ply  Stockinette topology is 3D printed using the Polyjet technology. Manual unraveling of the top row shows individual yarns and fibers. \textbf{c}, A $6 \times 6 \times 6$ volumetric knit showing an additional looping per stitch in the Pile direction. A periodic unit consists of many disjointed yarn segments. The scale bar is \SI{10}{\milli\metre}.}
    \label{fig:1}
\end{figure}

\section{Results}\label{sec2}
\subsection{Traditional vs. 3D printed knits}
First, we demonstrate that 3D printed knits can faithfully reproduce the mechanical signatures of their conventional counterparts. We therefore examine the construction of planar knitted fabrics—both traditional and 3D printed—to establish quantitative correspondence and to isolate the role of geometry in governing their response.

We begin by replicating the hierarchical structure of traditional knits from the loop topology to yarn arrangement and fiber count. With industrial knitting machines, a homogeneous knit fabric is constructed using a single contiguous yarn. Topologically, this yarn is arranged into multiple loops (or stitches) in series to form a row. At the end of a row, the yarn loops to begin the next row in reverse. In this next row, the loops are pulled through the loops in the previous row to create the knit fabric (Fig.~\ref{fig:1}a). The row and column directions are termed course (C) and wale (W) respectively (Fig.~\ref{fig:1}a).

To 3D print knits, we begin by mathematically defining the geometry of each loop. This enables implicit control of the topology of the knit, \textit{i.e.}, the manner in which the yarn loops itself. With such a description, we can directly generate and 3D print a volumetric representation of a knit using commercial technologies, such as inkjet 3D-printing~\cite{stratasysJ35Pro2023}. The resulting prints replicate the topology, geometry as well as the unraveling characteristics of a traditional knit (Fig.~\ref{fig:1}b).
Unlike industrial knitting machines however, 3D printing enables the exploration of patterns that were previously infeasible. By introducing the Pile direction, we demonstrate volumetric knit architectures that interloops both in plane and out-of-plane (Fig.~\ref{fig:1}c). The unit cell of which features a complex microstructure with multiple seemingly disjointed filaments.

\subsection{3D printing knit architecture}
\begin{figure}[b!]
    \centering
\includegraphics[width=0.5\textwidth]{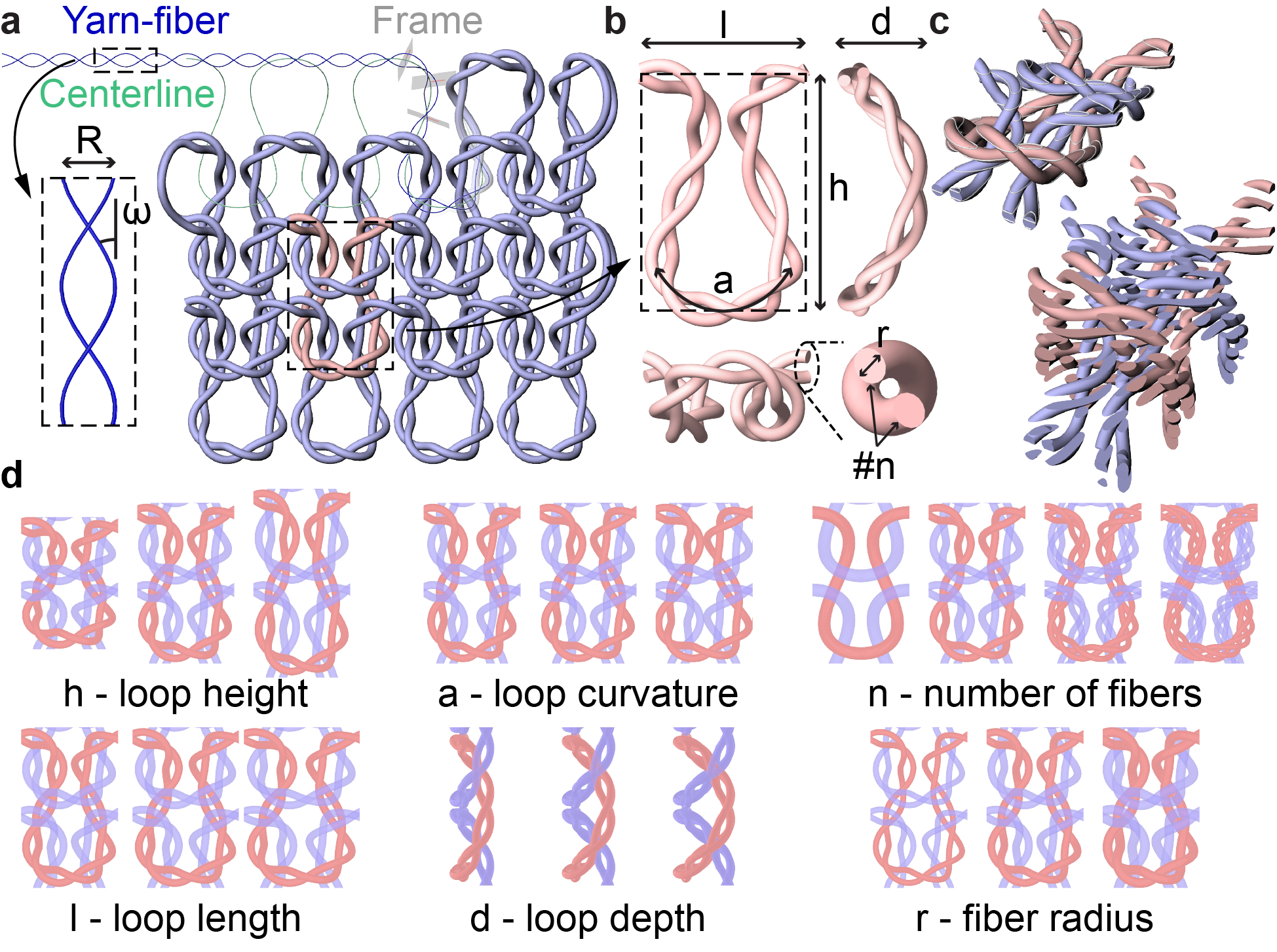}
    \caption{\textbf{Design and fabrication of 3D printed knits.} \textbf{a}, A centerline is defined, and a yarn-fiber spirals around the centerline based on its Frenet frames. A circular cross-section is assigned to each fiber to create a solid geometry. The yarn-fiber is parameterized by the separation distance and tilt angle.
    \textbf{b}, Parametric design space of the knit fabrics, featuring hierarchical architecture of fibers, yarns, and loop. \textbf{c}, Schematic of layer slicing and layer-by-layer 3D printing using the Polyjet. \textbf{d}, Influence of different geometric variables on the shape of the fabric, including loop height, length, depth and curvature, and yarn fiber number and radius.}
    \label{fig:2}
\end{figure}

Unlike industrial knitting machines where one only needs to specify the pattern of the fabric~\cite{singal2024programming}, 3D printing requires a geometrically valid closed triangulated surface mesh~\cite{pasquier2025fiber,wirth20233d}. We begin by defining the centerline of the knit. We then spiral the fiber bundles of the yarn around this centerline~\cite{crane2023SPM} (Fig.~\ref{fig:2}a).
As a canonical example, the centerline of a Stockinette stitch $\gamma(t)$ is geometrically described using an arc-length formulation as 

\begin{equation}
\gamma(t)=\left[\frac{l}{2\pi}\left(t+a\sin{2t}\right), h \cos{t}, d \cos{2t}\right],
\label{eq:parameterization}
\end{equation}

, where $t\in[0,2\pi]$ is the arc-length, $h, l, a, d$ are the height, length, curvature and depth of individual loops respectively~\cite{crane2023SPM} (See Fig.~\ref{fig:2}b, and SI for detailed derivations).

We then calculate the Frenet frames around this centerline to define the fiber bundles that spirals around (Fig.~\ref{fig:2}a). The fiber bundles are parametrized by the distance of separation of the fibers $R$, tilt angle $\omega$, fiber radius $r$ and count $n$  (Fig.~\ref{fig:2}b). Curved segments are added between adjacent rows in the Wale direction to maintain contiguity. The same knit and yarn geometry can be seen in a cotton knit of the same topology (Fig.~\ref{fig:1}a).

Using an inkjet 3D printing (Stratasys J35), a method where droplets of photopolymer are jetted onto the build platform and then cured with UV lamps to bond layers, we fabricated printed knits from the computational parametrization described above.
When the knit is printed in the as-knit configuration, the topological state of entanglement is inherently achieved so long as the triangulated surface mesh contains sufficient clearance such that the yarn does not fuse to itself during printing at crossing points (between loops and between rows). The layer-by-layer deposition along with a water-soluble sacrificial material ensures the fiber bundles remains distinct (Fig.~\ref{fig:1}c).

To examine the mechanical behavior of such structures, we systematically vary the geometric and material parameters. Specifically, we vary $h, l, a, d, n$ and $r$ which results in different loop geometries (Fig.~\ref{fig:1}d). Note that when increasing $n$, the radius $r$ is decreased to maintain total yarn cross sectional area, and to isolate the effect of $n$.
Leveraging multi-material printing, we experiment with two different materials, namely RGD8530-DM (shortened as RGD, $E=\SI{0.82}{\giga\pascal}$) and VeroUltraWhite (shortened as VUW, $E=\SI{2.0}{\giga\pascal}$~\cite{Stratasys2017DigitalABSPlus}) (see SI for material behavior). 

\subsection{Mechanics of planar knits}

Having shown that 3D printed knits are geometrically similar to traditional knits, we begin to explore their mechanical behavior. We subject the 3D printed knits (dimensions $L,H$) to cyclic equibiaxial strains up to 40\% in tension, and measure the reaction force in both the Course and Wale directions, $F_\mathrm{C}$ and $F_\mathrm{W}$. 
Using the overall sample dimensions, we calculate the effective strains and stresses $\varepsilon_\mathrm{eff}$ and $\sigma_\mathrm{eff}$ in both directions. Considering that knits are intended to be used as membranes, unit thickness is assumed.

\begin{figure}[ht!]
    \centering
\includegraphics[width=0.5\textwidth]{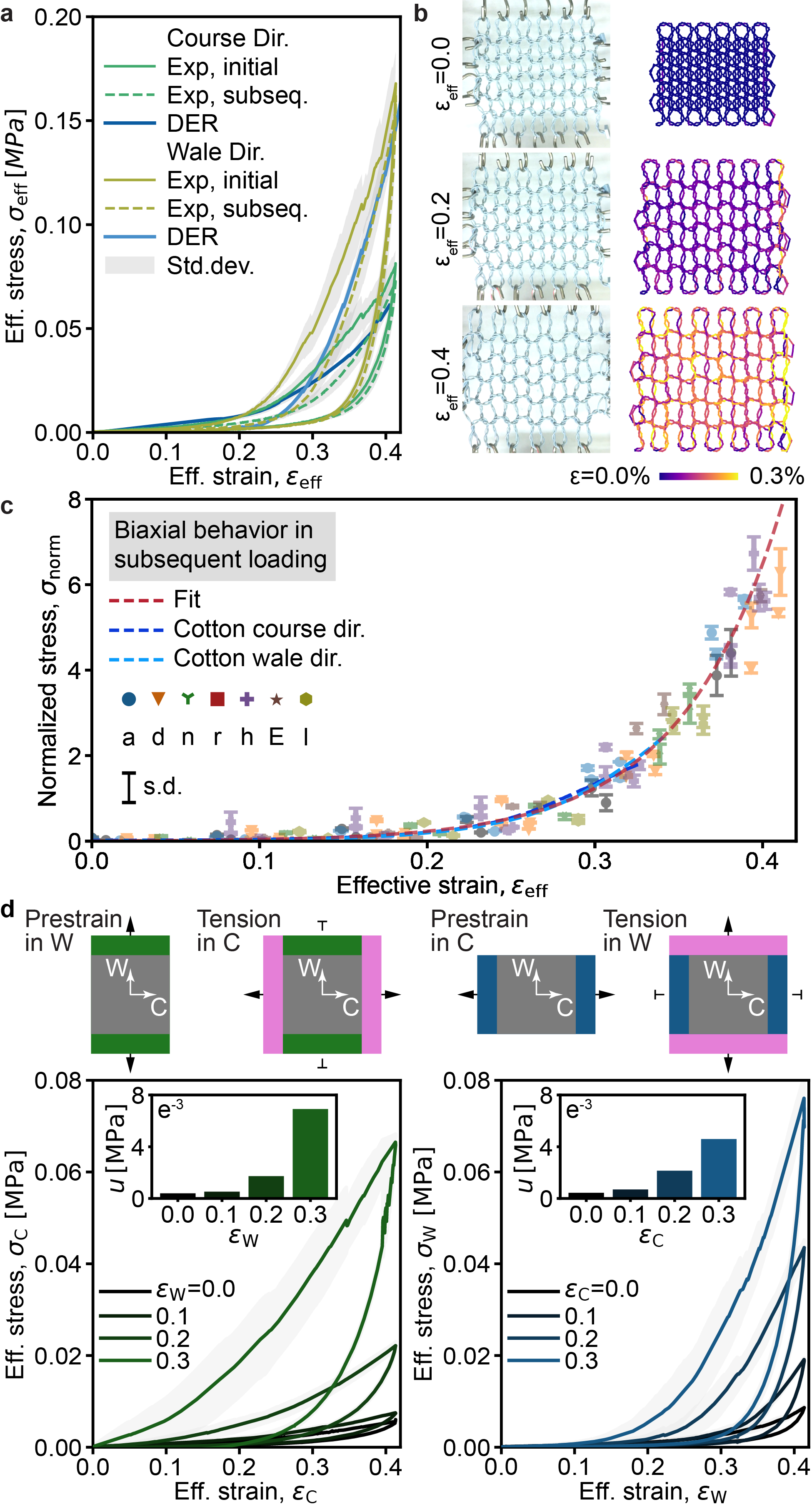}
    \caption{\textbf{Mechanical behavior of 3D printed knits.} \textbf{a}, Effective stress-strain plot of the knit in both Course (C) and Wale (W) directions of initial and subsequent stretching events. Both anisotropy and hysteresis are observed. \textbf{b}, Snapshots of equibiaxial stretching of a 6$\times$6 knit, both in experiments and using Discrete Elastic Rods (DER) simulations. \textbf{c}, Normalized stress-strain behaviors of knits printed with different geometric parameters, of a cotton fabric knit using a STOLL system, and of the exponential fit. \textbf{d}, Programmable stress-strain behaviors and strain energy dissipation characteristics in Course and Wale directions. In both cases, a pre-strain is imposed in one direction, then the orthogonal direction is loaded and unloaded.}
    \label{fig:3}
\end{figure}

The stress and strain plot of an arbitrarily chosen benchmark specimen exhibits an anisotropic, hysteric, exponential stress-strain behavior (Fig.~\ref{fig:3}a). Characteristic of knit fabrics, the initial loading exhibits a stiffer response as compared to subsequent loading~\cite{pasquier2025fiber}. The unloading curves of each experiment are identical. Similar to ``real'' fabrics, the Wale direction is stiffer than the Course, while pronounced hysteresis and dissipation are present. This is due to a combination of stitch dislocation, fiber re-arrangement and frictional contact as well as material viscoelasticity~\cite{jiang2017anisotropic}.

We exploit the fact that we have a complete geometric description of the knit to perform numerical simulations. Computation using volumetric finite elements remains expensive due to the complex frictional contact and sliding that occurs~\cite{pasquier2025fiber}.
Instead, a Discrete Elastic Rods (DER) based model is adopted~\cite{bergou2008discrete,yuksel2012stitch,sperl2022estimation}. The Incremental Potential Contact (IPC) toolkit is implemented for contact handling~\cite{li2020incremental,li2020codimensional,vidulis2023computational}. In addition to normal contact, tangential sliding is significant in knits where every stitch interacts with neighboring stitches. Therefore, Columbic tangential friction is introduced.
The model is tuned to predict subsequent loading events for practicality (Fig.~\ref{fig:3}b). In addition the global stress-strain behavior, the DER model allows the visualization of yarn-level strain. As the global strain increases biaxially, the curved loops first come into contact with neighboring loops, they then bend to accommodate the tension. At larger strains, since the same yarn travels across the entire fabric, each loops becomes rectangular and axial strain increases. This is distinctively different from uniaxial stretch experiments, where the orthogonal direction contracts to provide slack and the overall arc-length remains unchanged~\cite{poincloux2018geometry,crassous2024metastability}.

\begin{figure*}[ht!]
    \centering
\includegraphics[width=\textwidth]{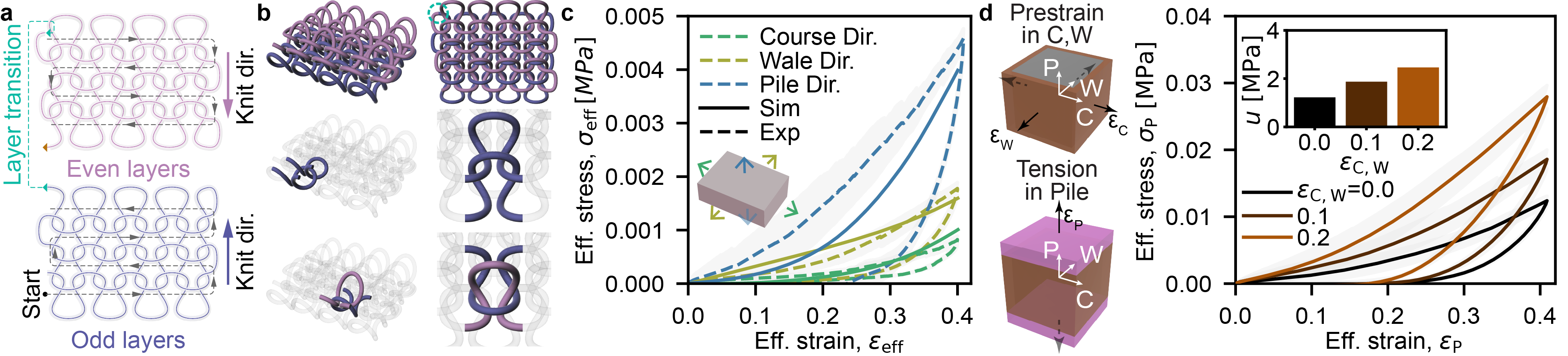}
    \caption{\textbf{3D printed volumetric knits} \textbf{a}, Topology of a volumetric stockinette pattern where even layers are knit in the opposite direction as the odd layers. \textbf{b}, When layered in the Pile direction, a 4$\times$4$\times$2 geometry is formed from a single continuous centerline. The fibers then spiral around this centerline to form the knit. Two distinct loops are present, the first interlaces neighboring rows in the Wale direction, the second is introduced to interlace neighboring layers in the Pile direction. \textbf{e}, Uniaxial stretch of a 6$\times$6$\times$6 knit in the three directions exhibit pronounced anisotropic and hysteresis. \textbf{d}, Programmable response in the Pile direction through equibiaxial strains imposed in both Course and Wale directions.} 
    \label{fig:4}
\end{figure*}

With the parametric sweep, we observe that the width $l$ and the height $h$ normalize the measured force and displacement to an effective stress $\sigma_\mathrm{eff}$ and strain $\varepsilon_\mathrm{eff}$. An increase in loop depth $d$ and loop curvature $a$, however, decrease the knit stiffness even as its areal density is increased. Conversely, increasing $n$ and $r$ result in stiffer behavior (see SI for the behavior of individual parametric variations). With these observations, we propose a dimensionless normalization factor,

\begin{equation}
\xi = v \frac{d^3 a^2}{E^2n^2r^4}
\label{eq:normalization}
\end{equation}

\noindent to derive a normalized stress measure $\sigma_\mathrm{norm}=\xi \sigma_\mathrm{eff}$ (Eq.~\ref{eq:normalization}) that accurately collapses the behaviors to a master curve (Fig.~\ref{fig:3}c). In $\xi$, in addition to the geometric parameters shown in Fig.~\ref{fig:2}d, $v$ accounts for the direction of loading with $v_\mathrm{C}=0.5$ and $v_\mathrm{W}=1.0$, and $E$ is the Young's Modulus of the print materials.

Using the same normalization with the biaxial behavior of a fabric (Stockinette, cotton) knit using an industrial knitting machine (Stoll CMS), we arrive at the same collapsed behavior, demonstrating that the mechanical behavior of machine-fabricated knits are qualitatively similar to the 3D printed counterparts (Fig.~\ref{fig:3}c).

We further propose an empirical relationship to map the effective strain to stress, $\sigma_\mathrm{eff} = \alpha e^{\beta \varepsilon_\mathrm{eff} }$. Where $\alpha=900$ and the empirical coefficient $\beta=16.50$ can best be explained as a topological entanglement constant. With this, we are able to directly predict the stress-strain behavior of a knit under arbitrary geometric parameterizations.

\subsection{Programmable behavior of knits}
The fact that 3D printed knits are all composed of a single yarn hints there exist coupled interactions between the orthogonal stretches of the fabric. We leverage this to demonstrate programmable mechanical behavior. Specifically, we seek to tune the stress-strain relationship in one direction by imposing a pre-strain in the other directions. 

We first impose a variable strain between 0.0 to 0.4 in either the Course or the Wale direction. Then we apply a prescribed strain in the other direction up to 0.4. The results show that we can predictably tune the stress output in both Course and Wale directions
(Fig.~\ref{fig:3}d). More pre-stretch in one of the directions increases both the stiffness and the strength of the orthogonal direction. 
Further, as the specimens recover to the initial position when unloaded, the hysteric behavior is repeatable and tunable with a larger pre-stretch leading to a larger hysteresis. 

The increase of Wale direction pre-strain increases the Course direction secant stiffness up to 0.4\% by an order of magnitude from 0.015 to \SI{0.166}{\mega\pascal}. Conversely, the pre-strain in the Course direction increases the Wale direction stiffness from 0.021 to \SI{0.190}{\mega\pascal}.
Correspondingly, the dissipated energy density increases from 0.39e-3 to \SI{6.9e-3}{\mega\pascal} and from 0.43e-3 to \SI{4.6e-3}{\mega\pascal} for the Course and Wale direction loading respectively, both representing an increase of over an order of magnitude. Consequently, such knits can be used as a damper to dissipate energy in a tunable manner.

\subsection{Entanglement towards volumetric knits}

We propose the design of a volumetric knit to showcase the ability for 3D printing to fabricate arbitrary filamentous entanglement. To form a volumetric knit, we introduce the Pile (P) direction in addition to the Course and Wale. On odd layers, the yarn traverses the fabric in the same manner as the planar knits. It loops from one end to the other on the odd rows, before doubling back on the even rows. Once the entire layer is knit, the yarn travels in the Pile direction to arrive at an even layer, where the knitting direction is reversed (Fig.~\ref{fig:4}a).  

In addition to the looping between neighboring rows, the trigonometry of the centerline formulation is modified to introduce a second form of interlooping that occurs between neighboring layers ~\cite{hirose2024solid}(Fig.~\ref{fig:4}b). Specifically, the loops are much taller such that they can loop with the loops above itself.
An additional connecting loop is introduced in every layer to connect the ends from one layer to the next. These changes are applied to the parameterization of the centerline, from which the fibers are computed through a new set of Frenet frames. The derivations are detailed in the SI.

With these small geometric changes, previously separated knit layers become topologically entangled during printing, yet the knit still maintain the contiguity of a single yarn. This knit structure can be interpreted as an architected material formed by periodically tessellation a single unit cell (Fig.~\ref{fig:1}c) even through the boundary planes of the unit cell feature disjointed filaments. As such, we conduct uniaxial tension experiments in all three directions. The results show anisotropy with the pile direction being the stiffest (Fig.~\ref{fig:4}c). As a multi-layered knit, however, the equibiaxial experiments along the Course and Wale directions show similar behavior as in the 2D case (See SI). Numerical modeling of the entire volumetric knit structure using DER accurately predicts the loading path of the stress-strain behavior (Fig.~\ref{fig:4}c).

Programmable mechanical behavior in the Pile direction is achieved by imposing pre-strains in both Course and Wale directions equi-biaxially. A custom triaxial loading frame composed of a biaxial and a uniaxial system is used to experimentally demonstrate the programmability. Equi-biaxial strains of up to 0.2 is imposed. The Pile direction is then tensioned until a strain of 0.4. The results show that the effective stiffness as well as the dissipated energy (area between the loading and unloading curves) in Pile direction can be programmed  (Fig.~\ref{fig:4}d).

\begin{figure}[ht!]
    \centering
\includegraphics[width=\textwidth]{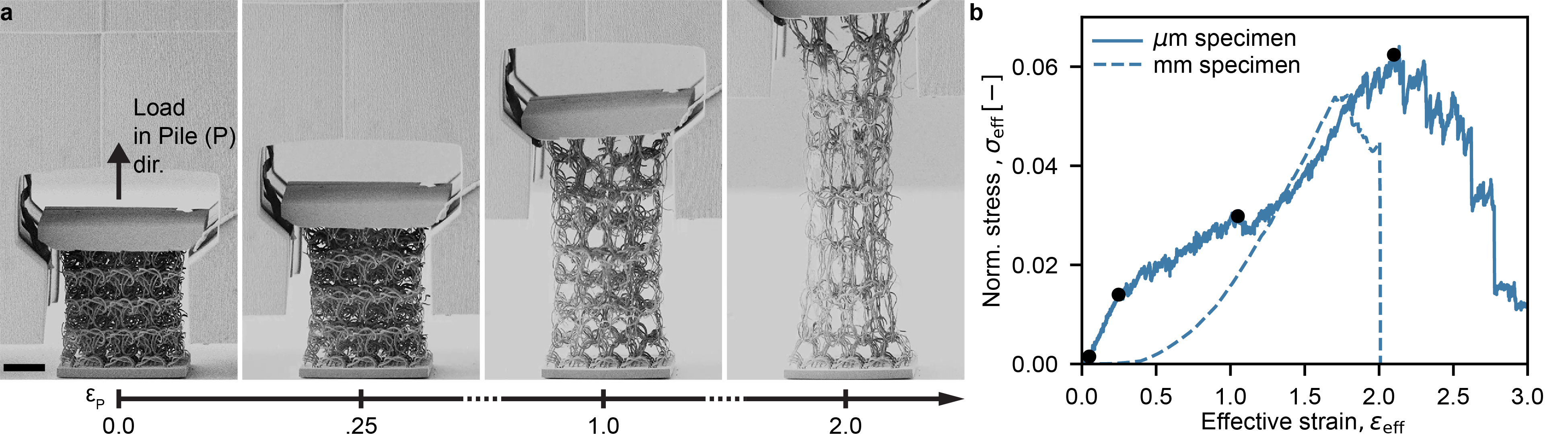}
    \caption{\textbf{Uniaxial tension of a microscopic volumetric knit.} \textbf{a}, A volumetric knit consisting of 6 loop stitches in each of the Course, Wale and Pile directions printed using the nanoscribe GT2. The scale bar is \SI{50}{\micro\metre}. \textbf{b}, Uniaxial tension experiments comparing the microscopic volumetric knit verses one shown in Fig.~\ref{fig:1}c up to rupture.}
    \label{fig:5}
\end{figure}

\subsection{Microscopic knit architecture}

To demonstrate the scale invariance of our design framework, we fabricate the same volumetric knit geometry at the micrometre scale using two-photon lithography (Nanoscribe GT2) with a loop size of approximately \SI{50}{\micro\metre}. The structures are printed in IP-Dip photoresist ($E=\SI{3.0}{\giga\pascal}$), with the top and bottom layers fused to load plates to facilitate mechanical testing at the micro-scale. Uniaxial tension experiments are performed \emph{in situ} in the pile direction inside a scanning electron microscope, using a silicon microgripper coupled to a displacement-controlled nanoindenter (Alemnis AG, see Methods)~\cite{surjadi2024double}. 

We then compared the response of the microscale structures with their macroscale counterparts printed with the inkjet process.
The results exhibit comparable deformation sequence at both scales (Fig.~\ref{fig:5}a).
By normalizing the stress response with respect to the material stiffness, the resulting behavior between the macro and the micro experiments are qualitatively similar. The initial stiffening in the micro-scale knit structure can be attributed to localized fusing between fibers. 
At a strain of approximately 1, these fusing points rupture and the remaining loading curve up to the ultimate strength follows a similar stiffness as the macroscale structure, and both reach approximately the same normalized ultimate tensile strength. While the brittle failure of the macroscale specimen can be attributed to the brittle nature of the inkjet constituent material, the similarities confirm that the governing mechanism—entanglement-mediated load transfer—is geometric rather than material.
Further, this correspondence across four orders of magnitude in length demonstrates that the geometry and physics of knitted architectures can be applied to tangible fabrics as well as, to our knowledge, the smallest knit ever fabricated.


\section{Methods}
\emph{Fabrication and characterization of large-scale specimens} The large-scale specimens studied in this manuscript are fabricated using the multi-material Polyjet technology (Stratasys J35). WSS150 is used as the water soluble sacrificial support material. All specimens are soaked in water for a period of 24 hours followed by a drying cycle of 24 hours prior to mechanical testing. 
The constituent material behavior is discussed in the SI.

Biaxial experiments are conducted using a custom stage described in detail in~\cite{pasquier2025fiber}. Each boundary loop is connected with two S-shaped hooks thatt slide on a metal pole. This enforces displacement in one direction while allowing nearly friction free sliding in the other. Uniaxial experiments use similar clamp setup in conjunction with a commercial testing machine (Instron 68SC-2). Triaxial tensile loading is accomplished by bringing the biaxial stage onto the uniaxial testing setup. All experiments were performed at a displacement speed of \SI{0.2}{\milli\metre\per\second}.

\emph{Fabrication and characterization of microscopic specimens}
All specimens were manufactured on silicon (Si) substrates using IP-Dip, an acrylate-based photoresist, via two-photon polymerization with a Photonic Professional GT2 system (Nanoscribe GmbH) using the 63x objective. A laser power of \SI{32.5}{\milli\watt} and a scan speed of \SI{10}{\milli\metre\per\second} were used for the knitted portion, while a laser power of \SI{25}{\milli\watt} and a scan speed of \SI{10}{\milli\metre\per\second} were used for the monolithic portion (that is, the tensile fixture atop the volumetric knit) to mitigate cavitation. A hatching and slicing distance of \SI{0.2}{\micro\metre} were used to fabricate the specimens. After printing, the samples were immersed in propylene glycol monomethyl ether acetate to remove uncured resin for approximately 5 hours. This was followed by a 10-minute rinse in isopropanol. The specimens were then dried using a critical point dryer (Autosamdri 931, Tousimis). Subsequently, the support structures were removed via plasma ashing in air for 30 minutes at \SI{100}{\watt} (see SI Fig. X). Finally, a 10-nanometer gold coating was applied via sputter coating (SCD 040, Balzers) to enable proper imaging during in situ mechanical tests.

To facilitate real-time observation of the deformation, uniaxial tension tests were performed inside an SEM (Gemini 450, ZEISS) using a custom tensile gripper attached to a nanoindenter (Alemnis AG). The gripper was operated in displacement-controlled mode, with all specimens subjected to a strain rate of approximately \SI{0.001}{\per\second}. Stress–strain data were obtained by normalizing the load–displacement measurements with the nominal cross-sectional area and the specimen height, respectively.

\emph{Numerical modeling using Discrete Elastic Rods}
A combination of Discrete Elastic Rods (DER) and the Incremental Potential Contact (IPC) toolkit is implemented for the simulation, developed based on the ElasticKnots model~\cite{vidulis2023computational}. DER uses principles from Kirchhoff’s theory and differential geometry to efficiently calculate the energy and equations of motion for flexible rods, and IPC acts as a customized nonlinear solver for efficient contact models~\cite{bergou2008discrete,li2020incremental,li2020codimensional}. While only normal contact force was considered for elastic knots, tangential force is significant in knitted structures where every stitch experiences frictional sliding with neighboring stitches. The tangential friction becomes significant with large strain, where the stitches interlock with each other with strong normal repulsion. Thus, we added the tangential friction component. We used the Coulomb friction model, where the friction force is proportional to the normal force and the direction of velocity. We introduced a friction coefficient that determines the magnitude of the friction force. Details are outlined in the SI.

\section{Discussion}
We have shown that 3D printing can successfully replicate both the geometry and mechanical behavior of traditional knit fabrics while overcoming the topological constraints imposed by industrial machines. Building on the inherent properties of looped yarns, we introduced volumetric knits in which the stitches extend and interloop across all three principal axes. Our approach demonstrates the rich mechanics within entangled material architectures. Future work leverage multi-material printing to tailor local stiffness, friction, or embed shape-memory and sensing elements for smart textiles, adaptive filters, and morphable reinforcements. Scaling down to nanoscale fibers further expands possibilities for tissue scaffolding, filtration, and multifunctional composites. Coupling the resilience and toughness of interlooped entanglement with additive manufacturing freedom thus paves the way for next-generation entangled material architectures.

\backmatter

\bmhead{Supplementary information}
The article is accompanied by one Supplementary Information file and four Supplementary videos. 

\bmhead{Acknowledgements}
The authors wish to thank P. Liu for her preliminary work on the biaxial testing stage and the Charm Lab at Stanford University (Dr. Okamura and Dr. Pasquier) for insightful discussions.

\section*{Declarations}

\begin{itemize}
\item Funding: NASA MIRO, NSF IIS/HCC, Haythornthwaite Foundation
\item Conflict of interest/Competing interests: Not applicable
\item Ethics approval and consent to participate: Not applicable
\item Consent for publication: Not applicable
\item Data availability: The data that support the findings of this study are present in the manuscript and/or in the Supplementary Information. Additional data are available from the corresponding author upon request.
\item Materials availability: Not applicable.
\item Code availability: The code that support the findings of this study are present in the manuscript and/or in the Supplementary Information. Additional code will be made available on a version control platform.
\item Author contribution: 
T.C. conceived this study.  B.C., T.C. designed the knit architectures. B.C., C.B., Y.W. fabricated samples, performed mechanical experiments and analyzed data. S.L. conducted DER simulations, L.X., J.U.S., C.M.P. conducted microscopic fabrication and experiments. C.M.P. and T.C. supervised the project. All authors contributed to the manuscript drafting.

\end{itemize}

\bibliography{ref.bib}

\end{document}